\documentclass[10pt, conference, compsocconf]{IEEEtran}

\usepackage{makeidx}
\usepackage{xspace}
\usepackage{listings}
\usepackage{url}
\usepackage{graphicx}
\usepackage{cite}
\usepackage{tipa}
\usepackage{multirow}
\usepackage{balance}
\usepackage{color}
\usepackage{colortbl}

\definecolor{mygray}{gray}{0.7}

\newcommand{\myparagraph}[1]{\subsubsection*{#1}}

\begin{document}

\newcommand{\etal}{et\,al.\xspace}
\newcommand{\ie}{\emph{i.\,e.},\ }
\newcommand{\eg}{\emph{e.\,g.},\ }
\newcommand{\sumup}[1]{\begin{center}\fbox{\begin{minipage}{0.9\columnwidth}\begin{center}#1\end{center}\end{minipage}}\end{center}}

\renewcommand{\paragraph}[1]{\vspace{1em}\par\noindent\textbf{#1.}}
\newcommand{\question}[1]{\vspace{1em}\par\noindent\textbf{RQ #1:}}

\title{Challenges of the Dynamic Detection\\ of
  Functionally Similar Code Fragments}

\author{\IEEEauthorblockN{Florian Deissenboeck, Lars Heinemann, Benjamin Hummel}
\IEEEauthorblockA{Technische Universit\"at M\"unchen, Germany\\
\{deissenb,heineman,hummelb\}@in.tum.de}
\and
\IEEEauthorblockN{Stefan Wagner}
\IEEEauthorblockA{University of Stuttgart, Germany\\
stefan.wagner@informatik.uni-stuttgart.de}
}

\maketitle

\begin{abstract}
Classic clone detection approaches are hardly capable of finding
redundant code that has been developed independently, \ie is not the
result of copy\&paste. To automatically detect such functionally
similar code of independent origin, we experimented with a dynamic
detection approach that applies random testing to selected chunks of
code similar to Jiang\&Su's approach. We found that such an approach
faces several limitations in its application to diverse Java systems.
This paper details on our insights regarding these challenges of
dynamic detection of functionally similar code fragments. Our findings
support a substantiated discussion on detection approaches and serve
as a starting point for future research.

\end{abstract}


\IEEEpeerreviewmaketitle

\section{Introduction\label{sec:intro}}
Research in software maintenance has shown that many programs contain
a significant amount of duplicated (cloned) code. Such cloned code is
considered harmful for two\linebreak reasons: (1)~multiple, possibly
unnecessary, duplicates of code increase maintenance
costs~\cite{koschke-2007_KoschkeR-clone_survey, roy-2007-clone_survey}
and, (2)~inconsistent changes to cloned code can create faults and,
hence, lead to incorrect program
behavior~\cite{2009_juergens_inconsistens_clones}. The negative impact
of clones on software maintenance is not due to copy\&paste but caused
by the semantic coupling of the clones. Hence, functionally similar
code, independent of its origin, suffers from the same problems clones
are known for.  In fact, the re-creation of existing functionality can
be seen as even more critical, since it is a missed reuse opportunity.

The manual identification of functionally similar code is infeasible
in practice due to the size of today's software systems. Tools are
required to automatically detect similar code. An earlier
study~\cite{juergens2010code} has shown that existing clone detection
tools are limited to finding duplicated code, \ie they are not capable
of finding redundant code that has been developed independently. As a
result, we do not know to what degree real-world software
systems contain similar code beyond the code clones that stem from
copy\&paste programming. Manual analysis of sample
projects~\cite{juergens2010code} as well as anecdotal evidence,
however, indicate that programs indeed contain many similarities not
caused by copy\&paste. Hence, we expected that tools for the automatic
detection of similar code could prove as beneficial for quality
assurance activities as the now widely-used clone detection tools.

While the equivalence of two programs is undecidable in general, a
straightforward approach to detect similar code relies on executing
candidate code fragments with random input data and comparing the
output values. This approach was pursued by
Jiang\&Su~\cite{jiang2009automatic} as well as us.  While they
successfully applied their approach for code in the Linux kernel, we
were unable to produce significant results using this seemingly simple
approach for diverse Java systems. As this differs from Jiang\&Su's
results, this paper details on our insights regarding challenges of
dynamic detection of functionally similar code in Java programs.

\myparagraph{Research Problem} 
Functional duplication in software systems causes a multitude of
problems for software maintenance. Although clone detection is a
viable approach to find copied code, existing tools are not capable of
finding similar code that was created
independently~\cite{juergens2010code}.
Jiang\&Su~\cite{jiang2009automatic} developed a dynamic approach to
identify functionally similar code in C systems and experienced a high
detection rate for the Linux kernel.  We implemented a similar
approach for detecting functionally similar code fragments in Java
systems. When analyzing five open-source Java systems, we got
considerably lower detection rates. We found this to be caused by
limitations of the approach when applied to Java systems. As a
consequence, it is unclear if the dynamic approach can be applied in
practice to detect functionally similar code fragments in
object-oriented systems.

\myparagraph{Contribution} 
While not a replication in the strict sense, this paper transfers
Jiang\&Su's work to object-oriented software implemented in Java. We
describe our implementation of the dynamic detection approach, report
on the detection results, and provide a detailed comparison of our
approach and results to Jiang\&Su's work.  Moreover, we discuss the
challenges of the dynamic detection approach for object-oriented
systems and thereby provide a basis for a substantiated discussion as
well as directions for further research.

\section{Terms \& Definitions}

\myparagraph{Simion}
We define a \emph{simion} as a functionally \underline{simi}lar
c\underline{o}de fragme\underline{n}t regarding I/O
behavior\cite{juergens2010code}.  Two fragments are simions if they
compute the same output for all input values.

\myparagraph{Chunk}
A \emph{chunk} is a code fragment that is compared for functional
similarity. It consists of a set of input parameters, a statement sequence,
and a set of output parameters. 

\myparagraph{Clone}
We define a \emph{clone} as a syntactically similar code fragment
typically resulting from copy\&paste and potential
modification. Several specific clone types were introduced that impose
constraints on the differences between the code
fragments\cite{koschke-2007_KoschkeR-clone_survey}.  \emph{Type-1
  clones} are clones that may differ in layout and comments.
\emph{Type-2 clones} may additionally differ in identifier names and
literal values. \emph{Type-3 clones} allow a certain amount of
statement changes, deletions or insertions. \emph{Type-4 clones} as
defined by \cite{roy-2007-clone_survey} are comparable to
simions. However, we do not use the term type-4 clones, since the term
``clone'' implies that one instance is derived from the other, which
is not necessarily the case for simions according to our
definition. In addition to the known clone
types we define for this paper \emph{type-1.5 clones} as type-1 clones
that may be subject to consistent variable renaming. 
The clone types form an inclusion hierarchy, \ie all type-2 clones are
also type-3 clones, but there are type-3 clones that are not type-2
(and analogously for the other types). While clones may be (and often
are) simions, even textually identical fragments of code may emit
different behavior because of the type binding implied by the
surrounding context.

\section{Dynamic Detection Approach}
\label{sec:detection}

Our approach for dynamically detecting functionally similar code
fragments follows in principle that of Jiang\&Su
\cite{jiang2009automatic}, \ie it is based on the fundamental
heuristic that two functionally similar code fragments will produce
the same output for the same randomly generated input. We
exclude clones from the simion detection, however, since these can be found
with existing clone detection tools.  The main difference to the
approach of Jiang\&Su is that our approach targets object-oriented
systems written in Java whereas they address C programs.  The
detection procedure can be divided in five principal phases that are
executed in a pipeline fashion. We implemented a prototype of this
pipeline based on our continuous quality assessment toolkit
ConQAT\footnote{\url{http://www.conqat.org}}.

\begin{figure}
\centering
\includegraphics[width=0.5\columnwidth]{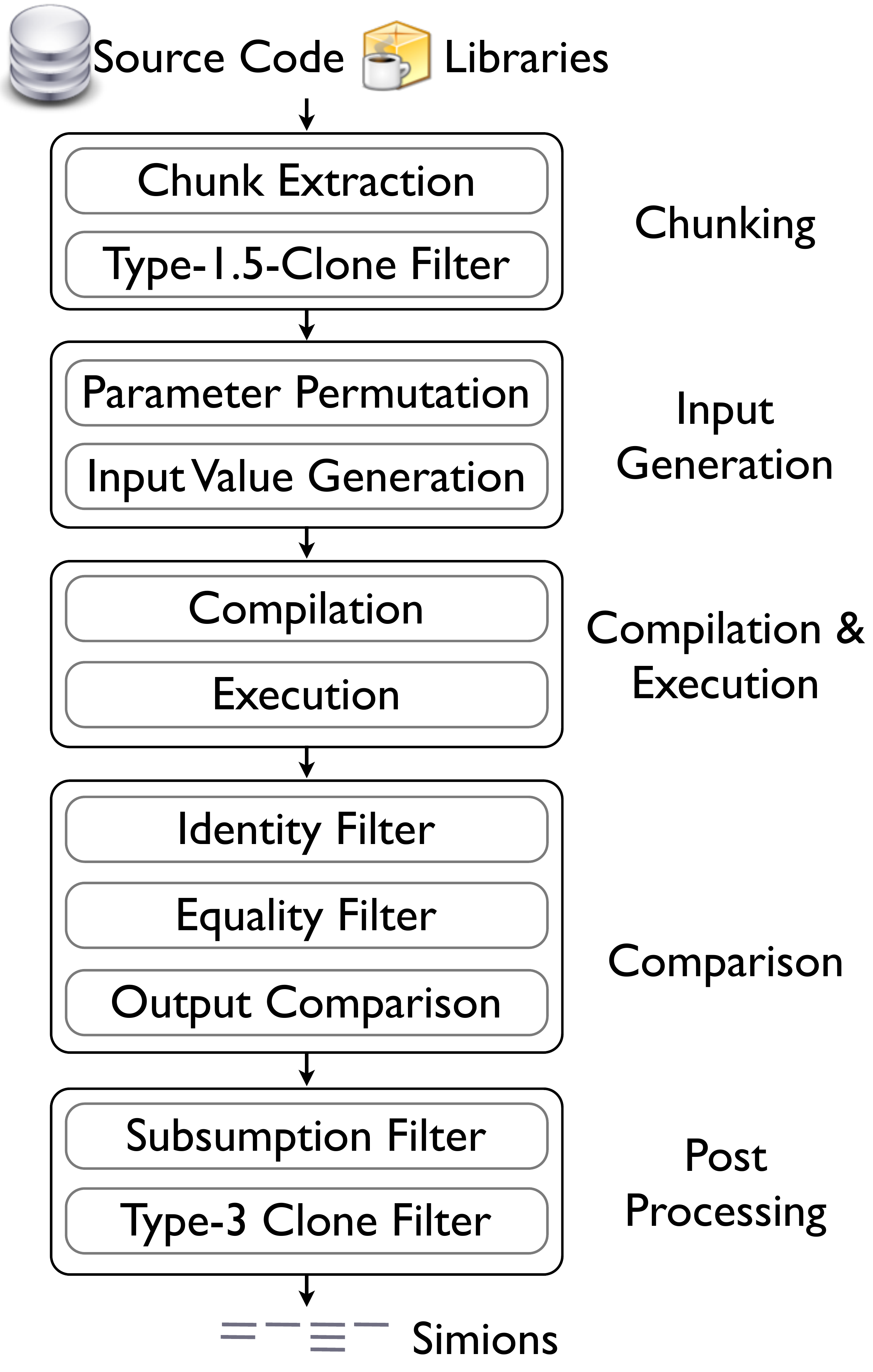}
\caption{Simion Detection Pipeline}
\label{fig:pipeline}
\end{figure}

Figure \ref{fig:pipeline} illustrates the detection pipeline with its
five phases, each consisting of several processing steps.  The input
of the pipeline is the source code of the analyzed projects and their
referenced libraries. The output is a set of equivalence classes of
functionally similar code fragments. The pipeline phases are detailed
in the following sections.

\subsection{Chunking}
\myparagraph{Chunk Extraction}
This step extracts a set of chunks from each
source code file, which are used as candidates for simion
detection. For Java systems, which consist of classes, it is a
challenge to extract those chunks, since a code fragment arbitrarily
cut out of the source files will almost certainly not represent a
compilable unit on its own. Therefore, we developed several strategies
for extracting chunks from Java classes.  The first challenge is to
determine what will be the input and output parameters of the
chunk. In a Java class, several types of variables occur. During the
chunk extraction we derive the input and output parameters from the
declared and referenced variables in the statements of the chunk with
the following heuristics:

\begin{itemize}
\item Referenced \emph{instance variables (non-static)} become input
  as well as output parameters.
\item Referenced \emph{class variables (static)} become neither input
  nor output variables, but are treated as a local variable as some of
  the called methods might rely on the value of these globals.
\item \emph{Local variables} with a scope that is nested in the
  statements of the chunk do not have to be considered as input or
  output variables. All other referenced local variables become output
  variables. Local variables that are referenced but not declared in
  the chunk statements also become input variables.
\item \emph{Method parameters} are treated in the same way as
  referenced but not declared local variables, \ie they become input
  and output parameters. They are also output parameters, as the code
  after the chunk might reference these values.
\end{itemize} 

Since we want to compare individual chunks for functional similarity,
we have to be able to compile their statement sequences and execute
them separately. We therefore process each statement sequence and
apply several transformations to obtain a static function with the
input and output parameters of the chunk. In certain cases, two chunks
are generated for the same statement sequence. This is used to provide
different input signatures for the same piece of code. For example,
code referencing the attributes of the class could either use an input
object of the class' type or inputs for each of the attributes,
leading to different chunk signatures.  There are cases when we
cannot create a compilable chunk at all. Examples include statement
sequences with branch statements (\eg \texttt{continue}) where the
target of the branch statement is not in the sequence or calls to
constructors of non-static inner classes.

Non-static methods that could be static since they do not reference
any non-static methods or fields are converted to static methods by
adding the static keyword to the method declaration. Thereby
potentially more chunks do not need an additional input parameter with
the type of the surrounding class and thus represent a more generally
reusable fragment.

In case that a statement sequence contains a return statement, the
resulting chunk gets a single output parameter with the type of the
return value of the associated method. The code with the return
statement is changed to a local variable declaration with the
corresponding initialization followed by a jump to an exit label.

The detection performance heavily depends on the number of
chunks compared for similarity. Since Jiang\&Su experienced performance
problems due to high numbers of chunks, we developed three different
chunk extraction strategies.

The \emph{sliding window} strategy extracts chunks by identifying all
possible statement subsequences that represent valid AST fragments
with a certain minimal length. Thus, it can cover cases where an
arbitrary statement sequence is functionally similar to another
one. However, the number of chunks is quadratic in the number of
statements of a method.

The \emph{intent-based} strategy utilizes the programmer's intent by
interpreting blank lines or comments in statement sequences as a
logical separator between functional units. It extracts one chunk that
includes all statements in a method and all chunks that can be formed
from the statement sequences separated by blank or comment lines.

The \emph{method-based} strategy utilizes structuring of the code
given by the methods and considers the statement sequences of all
methods in the code. This strategy has a slight variation for
determining the output parameters. If the method has a non-void return
value the chunk gets one output parameter with the return type of the
method. Otherwise the chunk gets one output parameter with the type of
the surrounding class.

\myparagraph{Type-1.5-Clone Filter} Although code clones are often
simions w.r.t. to our definition, they are not useful as they could be
more easily detected with a clone detection tool. The type-1.5-clone
filter discards chunks that are type-1.5 clones. This removes the
clones from the results and improves detection performance as these
chunks are not further processed.

\subsection{Input Generation}

\myparagraph{Parameter Permutation}
To cover cases where two code fragments compute the same function
but have different parameter ordering, we additionally apply a parameter
permutation. For each chunk we generate additional
chunks where all input parameters of the same type are permutated. To cope
with combinatorial explosion, we constrain the number of additionally
generated permutations to 25.

\myparagraph{Input Value Generation}

As Jiang\&Su, we employ a random testing approach for generating input
values. We generate code for constructing random values for the input
parameters of the chunks. Since the chunks must be executed on the
same inputs for comparison, the input generation uses a predefined set
of values for primitive parameters (all Java primitives, their wrapper
types, and String). Input values for parameters with composite types
are generated with a recursive algorithm. All constructors (if any) of
the type are used for creating instances. The algorithm is applied
recursively to the parameters of the constructors, \ie again, random
values are generated and passed as arguments to the constructor. In
case of 1-dimensional arrays and Java 5 collections where the
component type is known, we generate collections of several randomly
chosen but fixed sizes and apply the algorithm recursively for the
elements of the collection. For chunks with many parameters or
constructors with many arguments, a large number of inputs would be
generated by this approach.  Therefore, we constrain the 
number of input values generated for one chunk to 100.

\subsection{Compilation \& Execution}

\myparagraph{Compilation}
The \emph{compilation} step wraps the chunk code with the input
generation code and code for handling errors during execution and for
storing the values of the output parameters after execution. If an
exception occurred during the execution, a special output value
\emph{error} is used. Moreover, we generate code for handling
non-terminating chunks (\eg due to infinite loops), which stops
execution after a timeout of 1 second. The code is compiled in a
static method within a copy of the original class code and with all
compiled project code and libraries on the classpath. This means that
the statements in the chunk have access to all static methods and
fields from the original context as well as their dependencies.

\myparagraph{Execution}
After compilation, the chunks are executed in groups of at most 20 in
a separate Java Virtual Machine with a security manager
configured. This ensures that the chunk execution does not have
unwanted side effects (\eg deletion of files). The result of the
execution step is a list of execution data objects, which hold data
about the chunk and a list of input and output values of the chunk
execution.

\subsection{Comparison}

Preliminary experiments with the simion detection pipeline revealed
that the majority of identified simions is not relevant. A large
fraction are false positives, \ie our analysis identified them as
functionally similar although they are not. This is caused by the
random nature of the generated input data. Often two (or more) chunks
are executed with input data that \emph{triggers} only very specific
execution paths. With respect to these paths the chunks are similar
although they are not for reasonable input data. An example are two
string processing methods where one trims the string (deletes leading
and trailing whitespace) and one replaces characters with a unicode
code point above 127 with the unicode escape sequence used in Java. If
both chunks are executed for a string without leading or trailing
whitespace and without characters outside the ASCII range, they both
simply return the input string. As a result, they are identified as a
simion. We address both problems, clones and false positives, with
additional filter steps in our detection pipeline.

\myparagraph{Identity filter} The identity filter discards all chunks that 
implement the identity function, \ie for each input data set they return
the same output data set. This heuristic excludes chunks for which the randomly
generated input data is not capable of triggering \emph{interesting} execution
paths. 

\myparagraph{Equality filter} The equality filter discards all chunks that 
generate the same output data for all input data sets. The rationale
behind this is that the chunk execution is apparently independent of
the input data.  Again, this is caused by the inherently limited
quality of the randomly generated input data.  An example is a chunk
that has two input parameters: a string and an integer value $i$. If
$i$ is less than the length of the string (in characters), the chunk
returns a new string with the first $i$ characters of the input
string. If $i$ is greater than the length of the string, it returns
the empty string. As this dependency between the two parameters is
unknown to the input data generator it could possibly generate only
data sets where $i$ is greater than the length of the chosen
string. All executions of the chunk return the empty string.

\myparagraph{Output Comparison} The \emph{output comparison} step uses
the execution data objects to compare the chunks for functional
similarity.  Chunks that do not provide valid output data for at least
3 inputs (\ie either throw an exception or have a time-out), are
discarded at this step.  To make the comparison performant we use a
hash-based approach. For each chunk and each of its output variables,
it computes an MD5 digest for the comparison. This requires only
moderate space in memory even for large output data. This digest can
be thought of as a functional fingerprint of the chunk regarding a
projection to one of its output variables. To construct the MD5
digest, we transform each output object to a string representation and
append it to the MD5 digest. We use the XStream XML serialization
library\footnote{\url{http://xstream.codehaus.org}} to transform an
arbitrary object into a string with its XML serialization.  The MD5
digest of each chunk is used as a key into a hash map holding the
chunks. If two chunks have the same MD5 digest, we have identified a
pair of functionally similar code chunks. This is done for eliminating
the otherwise quadratic effort of comparing all chunks for equal MD5
digests. While hash collisions could lead to false positives, we
consider the comparison correct, since collisions are very unlikely in
practice.  The result of the output comparison step is a set of
equivalence classes of chunks with similar functionality.

\subsection{Post Processing}

\myparagraph{Subsumption filter} An additional filter discards simions
that are entirely covered by a larger simion (in terms of its length
and position in the source code). For example, if two methods are
identified to be simions, it is usually not worth reporting that parts
of them are also simions.

\myparagraph{Type-3 clone filter} 
Additionally, at the end of the pipeline a type-3-clone clone detector
is run to determine which of the simions could also be detected by a
type-3-clone-detector, \ie a clone detector that takes into account
insertion, modification, and deletion of a certain amount of
statements. The filter calculates the statement-level edit-distance
between chunks and is configured to filter all chunks with an
edit-distance less or equal to 5. Both clone filters are implemented
with ConQAT's clone detection
algorithm~\cite{2009_juergens_clonedetective}. We cannot filter type-3
clones earlier, as they are not guaranteed to be functionally
equivalent. Thus, it is unclear which of the instances should be
filtered as each could be a potential simion of another chunk.

\section{Study Design}

\subsection{Research Questions}

\question{1} How large is the simion detection problem?

As a first step, we are interested to characterize the problem we want
to solve. We look at the number of chunks extracted by our approach.

\question{2} How do technical challenges affect the detection?

To compare the functionality of two chunks, we need to transform them
into executable code pieces and generate useful test cases. This
provokes a number of technical challenges that affect the detection
approach.  This includes the generation of input values for the
chunks, which need to be meaningful to trigger interesting
functionality. Furthermore, the generation of useful test data for
project-specific data types and the emulation of certain operations
used by the chunks such as file I/O or GUI events is challenging.
Finally, even after overcoming these limitations, there can still be
problems that prevent the compilation of the extracted chunk. We
investigate how many chunks need to be disregarded during detection
because of these challenges.

\question{3} How effective is our approach in detecting simions?

We ask if the approach is able to detect significant amounts of
functionally similar code. Moreover, as we investigated the technical
challenges in RQ~2, we are interested in the share of the code of
real-world systems that we are able to analyze. Finally, we want to
know how many simions we can find in realistic systems.

\subsection{Data Collection Procedure}

In Section~\ref{sec:detection}, we described three
different chunking strategies. The general idea is to run a complete simion
detection with all strategies on a set of software systems to collect the needed data for
answering the research questions. For practical reasons, however, we cannot
perform a complete detection using the sliding window chunking strategy,
because it creates far more chunks than are feasible to analyze. Therefore,
we collect data using that strategy only for RQ1.

For RQ 1 and RQ 2, we employ no filters, since we are only interested in
determining the difficulty of the problem and technical challenges regardless of
the precision of the results. For RQ 3, where we are interested in the amount
of detected simions, we use all filters.

\myparagraph{RQ 1}
ConQAT writes the total number of extracted chunks into its log file,
which answers RQ~1. We use separate configurations
for the different chunking strategies.

\myparagraph{RQ 2}
To answer RQ 2, we use two different configurations.  The first one is similar to
RQ~1, using a specific statistics processor for collecting the required data.  Our
input generator logs the number of chunks for which no input could be 
generated.  Additionally, our configuration determines and counts the
types of the input parameters and aggregates these values. 

Another part of the configuration counts the number of chunks that
contain calls to I/O, networking, SQL, or UI code.  We cannot execute
chunks containing such calls successfully, as the expected files,
network peers, or databases are not available, or the required UI
initialization was not performed.  We identify calls to these groups
by the package the corresponding class resides in, \eg a call to
\emph{java.io.File.canRead()} would be counted as I/O. These packages
also contain methods that can be safely called even without the
correct environment being set up (such as methods from
\emph{java.io.StringReader}), so we expect to slightly overestimate
these numbers. On the other hand, we only count methods that are
directly called from the chunk. Methods that are called indirectly
(from other methods called from the chunk) are not included in these
numbers. However, as we are only interested in the magnitude of this
problem, we consider this heuristic sufficient.

The second configuration is a slightly simplified detection pipeline,
that uses the approach described in Section~\ref{sec:detection} to
generate code and tries to compile the chunks. Statistics on the
number of chunks that could not be compiled are reported.  For both
configurations, we disabled the type-1.5 clone filter and the
permutation step, as these distort the statistics slightly.

\myparagraph{RQ~3}
For the last research question we utilize the full simion detection
pipeline to count the number of simions detected by our implementation.
Our code is instrumented to report the number of chunks lost at the steps
of the pipeline.

\subsection{Analysis Procedure}

\myparagraph{RQ 1}
For the size of the problem, we report the total number of chunks per
chunking strategy to show the order of magnitude. To make the numbers
more comparable and to allow an estimate for the simion detection in
systems of other sizes, we give the number relative to the lines of
code (SLOC) and calculate the mean value.

\myparagraph{RQ 2}
We show the relative distribution and calculate the mean per strategy for each
of the following metrics, which characterize different technical challenges. We
give the values for all relative metrics rounded to full percentages.
First, we analyze the difficulty of generating inputs by two metrics. One is the 
number of chunks for whose input parameters we cannot generate values.
The other is the number of
inputs of project-specific data types, because it is especially 
hard to generate meaningful input for them.
Second, for the execution there are certain types of methods that are hard to emulate
during random testing. We analyze the number and share of calls to I/O,
network, SQL, and UI.
Third, the chunks need to be compiled to be executed. Hence, we
investigate the fraction of chunks that cannot be compiled. For these
challenges, we add qualitative, manual analysis of the chunks that cannot
be further used in the detection approach to get more insights into the
reasons.

\myparagraph{RQ~3}
We analyze two different types of study objects. The first type is a
set of programs of which we know that they have to exhibit similar
functionality, because they were produced according to the same
specification. These study objects show whether the detection approach
works in principle. We expect to get at least as many simions reported
as there are implementations of the specification. The second type
of study objects are real-world, large systems of which we do not
know beforehand of any simions. We show the change in chunks
during the execution of the detection pipeline in absolute and relative
terms. We round the relative values to percentages with two positions 
after the decimal point to be able to differentiate small results.

\subsection{Study Objects}

We chose two different types of study objects: (1) a large number of
Java programs that are functionally similar and (2) a set of
real-world Java systems.  The study objects of type 1 are small
programs, so that they are easy to analyze and they are built
according to the same specification so that we can be sure that they
exhibit largely the same functionality. We selected a set of programs,
we used in an earlier study~\cite{juergens2010code}.  We will refer to
it as ``Info1''.  They are implementations of a specification about an
e-mail address validator by computer science undergraduate
students. We only include programs passing a defined test suite to
ensure a certain similarity in functionality. This results in 109
programs with a size ranging from 8 to 55 statements.

The second type of study objects represents real Java systems to show
realistic measurements for chunks and simions.  As selection criteria,
we chose systems that cover a broad range of application domains,
sizes, and functionalities. Furthermore, we chose systems we are
already familiar with to support the interpretation of the
results. The selection resulted in the five open source Java systems
that represent libraries, GUI applications, and servers, shown in
Table~\ref{tab:projects} together with their size in SLOC
(\emph{source lines of code}, \ie the number of non-blank and
non-comment lines).

\begin{table}
\caption{Size of Open Source Study Objects}
\label{tab:projects}
\centering
\begin{tabular}{|l|r|} 
\hline
\textbf{Project} & \textbf{SLOC}\\
\hline
\hline
Commons Lang & 17,504\\
\hline
Freemind & 51,762\\
\hline
Jabref & 74,586\\
\hline
Jetty & 29,800\\
\hline
JHotDraw & 78,902\\
\hline
\hline
\textbf{Overall} & \textbf{252,554} \\ \hline
\end{tabular}
\end{table}

\section{Results}

\subsection{RQ 1: Problem Size}

Table~\ref{tab:number_of_chunks} shows the absolute and relative
numbers of chunks extracted for each study object and the different
chunking strategies. The sliding window strategy extracts the highest
number of chunks with up to 2.68 chunks per SLOC and 1.44 chunks/SLOC
on average.  The intent-based strategy creates less chunks with at
most 0.40 chunks/SLOC and 0.25 chunks/SLOC on average. The smallest
number of chunks is extracted using the method-based strategy. It
creates at most 0.09 chunks/SLOC and 0.05 chunks/SLOC on average.

Overall, the number of chunks is large, especially for the sliding
window strategy.  The detection approach needs to be able to cope with
several thousand chunks.

\begin{table*}
\caption{Total Number of Chunks for Different Extraction Strategies}
\label{tab:number_of_chunks}
\centering
\begin{tabular}{|l|r|r|r|r|r|r|} 
\hline
\textbf{Object} & \multicolumn{2}{|c|}{\textbf{Sl. Win.}} & \multicolumn{2}{|c|}{\textbf{Intent}} & \multicolumn{2}{|c|}{\textbf{Method}}\\
\hline
& Total & per SLOC & Total & per SLOC & Total & per SLOC\\
\hline
\hline
Commons Lang & 7,940 & 0.45 & 1,843 & 0.11 & 1,538 & 0.09\\
\hline
Freemind & 80,816 & 1.56 & 20,632 &0.40 & 1,984 & 0.04\\
\hline
Jabref & 133,556 & 1.79 & 21,388 & 0.27 &  2,085 & 0.03\\
\hline
Jetty & 22,006 & 0.74 & 7,713 & 0.26 & 1,457 & 0.05\\
\hline
JHotDraw & 211,283 & 2.68 & 16,221 & 0.21 & 2,813 & 0.04\\
\hline
\hline
\textbf{Mean} & -- & 1.44 & -- & 0.25 & -- & 0.05\\
\hline
\end{tabular}
\end{table*}

\subsection{RQ 2: Technical Challenges}

The first pair of columns in Table~\ref{tab:rq2} shows the fraction of
chunks for which the approach cannot construct input values. The two
main cases where no input can be generated are chunk parameters that
refer to (1) an interface or abstract class and (2) a collection with
an unknown component type\footnote{This second issue especially
  applies to code that is not targeted at version 1.5 or above of Java
  which supports Generics and thereby allows to specify the component
  type within the declaration of a collection type.}. In the first
case it is unclear which implementation should be chosen to obtain an
object that implements the interface.  In the second case we do not
know what type of objects to put in the collection. For all systems
the fraction for which no input can be generated is higher for the
intent-based chunking strategy compared to the method-based strategy.
In case of \emph{Freemind} and the intent-based chunking strategy for
as much as 94\% of the chunks no input could be generated. An
investigation revealed that \emph{Freemind} uses untyped
Collections. Except for \emph{Commons Lang}, where primitive types are
dominant, input generation failed for more than 30\% of the chunks
with both chunking strategies.

\begin{table}
  \caption{Fraction of chunks with technical challenges}
\label{tab:rq2}
\centering
\begin{tabular}{|l|r|r|r|r|r|r|r|} 
\hline
& \multicolumn{2}{c|}{No Input} & \multicolumn{2}{c|}{Proj. Specific} & \multicolumn{2}{c|}{Specific Env.} \\
\textbf{Project} & \textbf{Meth.} & \textbf{Int.} & \textbf{Meth.} & \textbf{Int.} & \textbf{Meth.} & \textbf{Int.}\\
\hline
\hline
Commons Lang & 20\%& 37\% & 60\%& 60\% & 1\% & 5\%\\
\hline
Freemind & 65\%& 94\% & 79\%& 75\% & 11\% & 27\% \\
\hline
Jabref & 32\%& 41\% & 61\%& 81\% & 17\% & 44\%\\
\hline
Jetty & 36\%& 66\% & 72\%& 87\% & 7\% & 15\% \\
\hline
JHotDraw & 43\%& 55\% & 76\%& 86\% & 24\% & 60\% \\
\hline
\end{tabular}
\end{table}

We determined how many chunks have parameters (either input or output)
of project-specific data types. The results are shown in the second
pair of columns in Table~\ref{tab:rq2}.  A
considerable fraction of the chunks (60--87\%) refer to
project-specific data types. With our approach, these would not
qualify as candidates for cross-project simions, since the other
project would not have the same data types.

\begin{figure}
\centering
\includegraphics[width=.49\columnwidth]{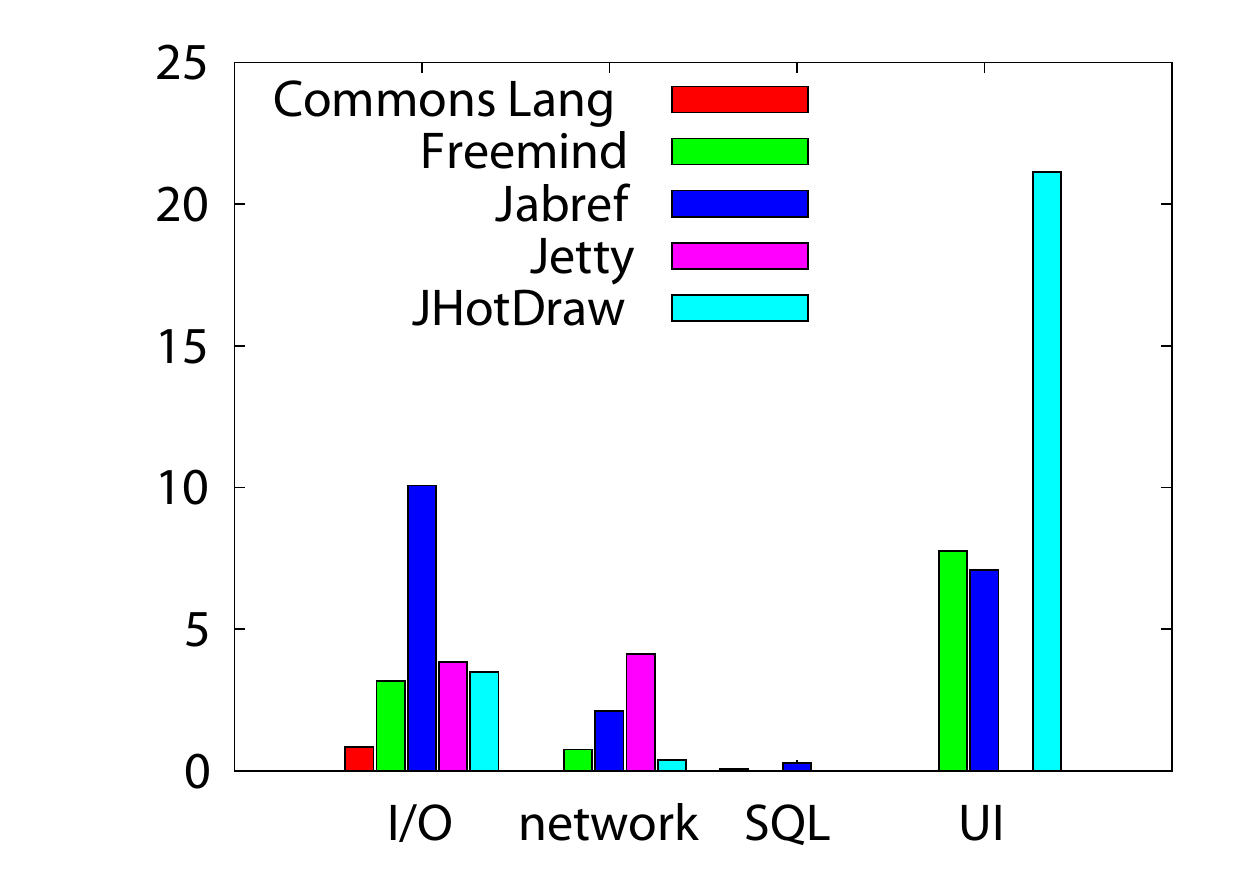}
\includegraphics[width=.49\columnwidth]{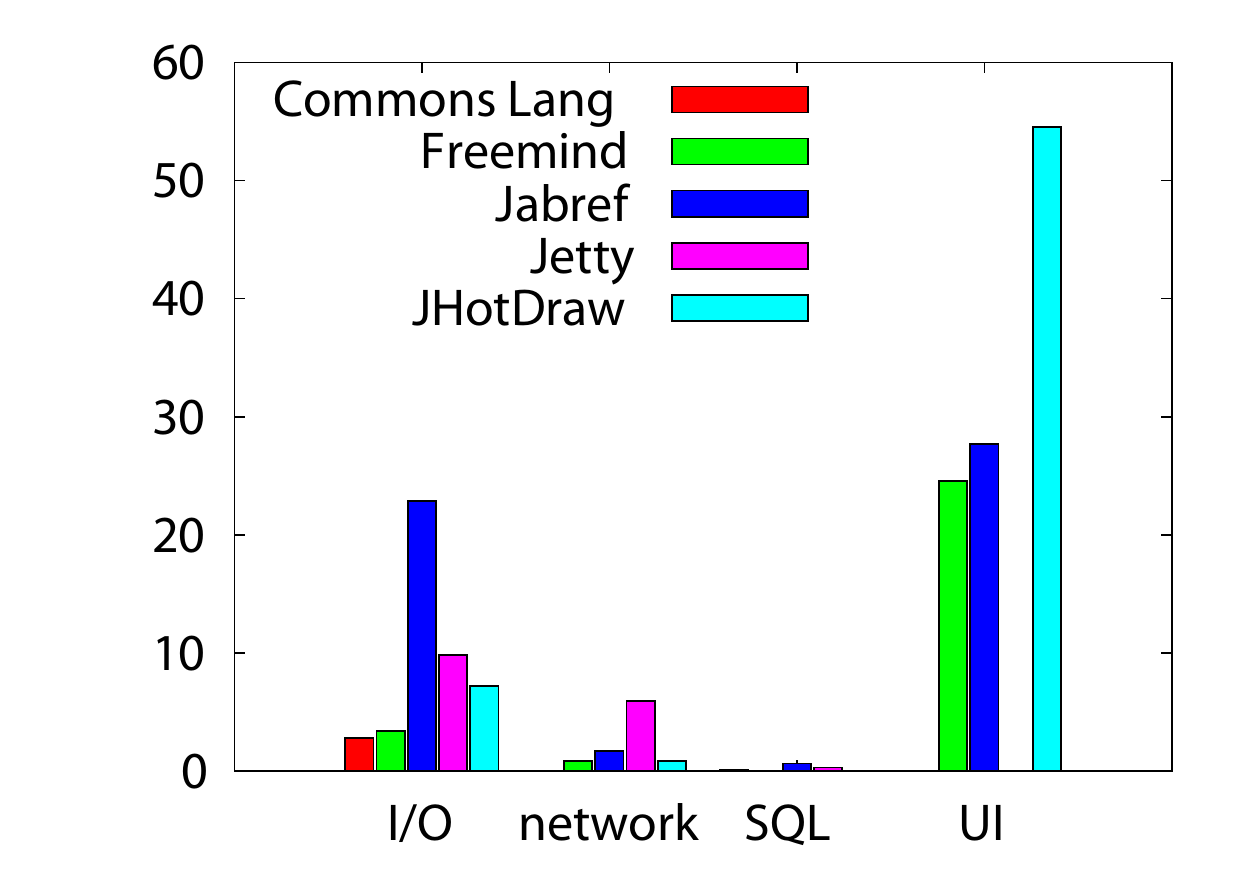}
\caption{Fraction of chunks calling complex methods(\%) for method
  chunking (left) and intent chunking (right)}
\label{fig:chunkCalls}
\end{figure}

To estimate the use of methods that require a specific environment, we
determined for each chunk whether it contains direct calls to methods
from one of the groups I/O, networking, SQL, or UI. The relative
numbers of chunks containing calls to a category are shown in
Figure~\ref{fig:chunkCalls} for each of the categories. It is not
surprising, that the numbers depend on the application. The library
\emph{Commons Lang} has only a couple of calls to I/O code, the HTTP
server \emph{Jetty} has the highest number of calls to networking
code, and the drawing tool \emph{JHotDraw} dominates the UI group. The
most common groups are UI (if the application has a user interface at
all), followed by I/O.

In the execution step, a chunk containing methods from at least one of
these groups is likely to fail as the expected environment is not
provided.  The last pair of columns in Table~\ref{tab:rq2} lists how
many chunks are affected by such methods for both of the chunking
strategies.  Overall, as many as 24\% of the methods can be affected
(\emph{JHotDraw}), or for the intent-based chunking strategy more than 60\%
of the chunks, thus having a significant impact on the number of
chunks we can process by our dynamic approach. Interestingly, the
relative numbers for the intent-based strategy are higher than for the
method-based strategy in all cases. This suggests, that the methods
containing I/O or UI code are typically longer than the remaining
methods and thus produce more chunks.

\begin{table}
  \centering  
  \caption{Compileable Chunks}
  \label{tab:compile}
    \begin{tabular}{|l|r|r|}
    \hline
    \textbf{Project} & \textbf{Method} & \textbf{Intent} \\
    \hline\hline
    Commons Lang & 96\% & 97\% \\\hline
    Freemind & 92\% & 99\% \\\hline
    JabRef & 90\% & 82\% \\\hline
    Jetty & 83\% & 88\% \\\hline
    JHotDraw & 93\% & 84\% \\\hline
    \end{tabular}
\end{table}

Finally, we checked how many of the chunks we were able to make
compilable by providing a suitable context. The relative number of chunks
we could make compilable of those for which at least one input could be
generated is shown in Table~\ref{tab:compile}. These numbers do not
indicate a principal limitation, as each of the chunks we extracted is a
valid subtree of the AST and thus can be executed in a suitable context.
They rather document limitations in our tool. An inspection of the
problematic chunks revealed weaknesses in chunks dealing with generic data
types, anonymous inner classes, method local classes, and combinations
thereof. Still, we are able to automatically generate a context for at
least 82\% in all cases and up to 99\% for \emph{Freemind}. Clearly, these numbers
could be improved by using more advanced algorithms for generating the
context for a chunk. The other results from RQ~2 and those of RQ~3
presented next suggest, however, that the chunks we lost as we cannot make them
compile is not the main bottleneck of the detection pipeline.

\subsection{RQ 3: Simions}

\begin{table*}
  \centering\footnotesize
  \caption{Analysis results for the method-based chunking strategy}  
    \begin{tabular}{|l|rr|rr|rr|rr|rr|>{\columncolor{mygray}}r>{\columncolor{mygray}}r|}
    \hline
          & \multicolumn{2}{c|}{\textbf{Comm. Lang}} & \multicolumn{2}{c|}{\textbf{Freemind}} & \multicolumn{2}{c|}{\textbf{JabRef}} & \multicolumn{2}{c|}{\textbf{Jetty}} & \multicolumn{2}{c|}{\textbf{JHotDraw}} & \multicolumn{2}{>{\columncolor{mygray}}c|}{\textbf{Info1}} \\
          & Abs.  & Rel.\,[\%]  & Abs.  & Rel.\,[\%]  & Abs.  & Rel.\,[\%]  & Abs.  & Rel.\,[\%]  & Abs.  & Rel.\,[\%]  & Abs.  & Rel.\,[\%] \\\hline\hline
    Chunk Extr. & 1,538 & 100.00 & 1,984 & 100.00 & 2,085 & 100.00 & 1,457 & 100.00 & 2,813 & 100.00 & 240   & 100.00 \\\hline
    Type-1.5-Clone & 1,100 & 71.52 & 1,541 & 77.67 & 1,647 & 78.99 & 1,034
    & 70.97 & 1,936 & 68.82 & 230   & 95.83 \\\hline Permutation & 1,601 & 104.10 & 1,916 & 96.57 & 2,245 & 107.67 & 1,255 & 86.14 & 3,472 & 123.43 & 231   & 96.25 \\\hline
    Input Gen. & 1,265 & 82.25 & 643   & 32.41 & 1,542 & 73.96 & 742   & 50.93 & 1,879 & 66.80 & 134   & 55.83 \\\hline
    Compilation & 1,215 & 79.00 & 530   & 26.71 & 1,324 & 63.50 & 568   & 38.98 & 1,586 & 56.38 & 133   & 55.42 \\\hline
    Execution & 1,066 & 69.31 & 189   & 9.53 & 621   & 29.78 & 313   & 21.48 & 660   & 23.46 & 133   & 55.42 \\\hline
    Identity & 1,066 & 69.31 & 189   & 9.53 & 621   & 29.78 & 313   & 21.48 & 660   & 23.46 & 133   & 55.42 \\\hline
    Equality & 947   & 61.57 & 165   & 8.32 & 522   & 25.04 & 178   & 12.22 & 579   & 20.58 & 133   & 55.42 \\\hline
    Comparison & 108   & 7.02 & 15    & 0.76 & 94    & 4.51 & 19    & 1.30 & 34    & 1.21 & 105   & 43.75 \\\hline
    Subsumption & 108   & 7.02 & 13    & 0.66 & 90    & 4.32 & 19    & 1.30 & 30    & 1.07 & 105   & 43.75\\
    \hline\hline
    Type-3-Clone & 54    & 3.51 & 11    & 0.55 & 55    & 2.64 & 15    & 1.03 & 18    & 0.64 & 105   & 43.75\\
    \hline
    \end{tabular}
  \label{tab:overal_mb}
\end{table*}

\begin{table*}
  \centering\footnotesize
  \caption{Analysis results for the intent-based chunking strategy}
    \begin{tabular}{|l|rr|rr|rr|rr|rr|>{\columncolor{mygray}}r>{\columncolor{mygray}}r|}
    \hline
          & \multicolumn{2}{c|}{\textbf{Comm. Lang}} & \multicolumn{2}{c|}{\textbf{Freemind}} & \multicolumn{2}{c|}{\textbf{JabRef}} & \multicolumn{2}{c|}{\textbf{Jetty}} & \multicolumn{2}{c|}{\textbf{JHotDraw}} & \multicolumn{2}{>{\columncolor{mygray}}c|}{\textbf{Info1}} \\
          & Abs. & Rel.\,[\%]  & Abs. & Rel.\,[\%]  & Abs. & Rel.\,[\%]  & Abs. & Rel.\,[\%]  & Abs. & Rel.\,[\%]  & Abs. & Rel.\,[\%] \\\hline\hline
    Chunk Extr. & 1,843 & 100.00 & 20,632 & 100.00 & 21,388 & 100.00 & 7,713 & 100.00 & 16,221 & 100.00 & 1,969 & 100.00 \\\hline
    Type-1.5-Clone & 1,598 & 86.71 & 20,586 & 99.78 & 20,794 & 97.22 &
    7,527 & 97.59 & 14,643 & 90.27 & 1,938 & 98.43 \\\hline Permutation & 4,690 & 254.48 & 26,705 & 129.43 & 61,836 & 289.12 & 17,131 & 222.11 & 51,683 & 318.62 & 3,259 & 165.52 \\\hline
    Input Gen. & 2,823 & 153.17 & 2,858 & 13.85 & 33,559 & 156.91 & 5,301 & 68.73 & 26,347 & 162.43 & 1,910 & 97.00 \\\hline
    Compilation & 2,772 & 150.41 & 2,432 & 11.79 & 22,837 & 106.77 & 4,227 & 54.80 & 16,527 & 101.89 & 1,883 & 95.63 \\\hline
    Execution & 2,399 & 130.17 & 556   & 2.69 & 3,998 & 18.69 & 2,173 & 28.17 & 9,887 & 60.95 & 1,841 & 93.50 \\\hline
    Identity & 2,298 & 124.69 & 424   & 2.06 & 3,480 & 16.27 & 1,734 & 22.48 & 9,352 & 57.65 & 1,825 & 92.69 \\\hline
    Equality & 2,014 & 109.28 & 388   & 1.88 & 2,818 & 13.18 & 1,392 & 18.05 & 8,694 & 53.60 & 1,730 & 87.86 \\\hline
    Comparison & 85    & 4.61 & 4     & 0.02 & 232   & 1.08 & 36    & 0.47 & 131   & 0.81 & 752   & 38.19 \\\hline
    Subsumption & 77    & 4.18 & 4     & 0.02 & 196   & 0.92 & 32    & 0.41 & 67    & 0.41 & 420   & 21.33 \\
    \hline\hline
    Type-3-Clone & 59    & 3.20 & 2     & 0.01 & 159   & 0.74 & 28    & 0.36 & 46    & 0.28 & 418   & 21.23 \\
    \hline
    \end{tabular}
  \label{tab:overal_ib}
\end{table*}

Tables~\ref{tab:overal_mb} and \ref{tab:overal_ib} summarize the analysis
results for all study objects discussed previously plus the additional
``Info1'' system (gray column). For each study object the table depicts
all pipeline steps along with the number of resulting chunks \emph{after}
the step's execution (column ``Abs.''). Additionally, column ``Rel.'' shows
the relative number of chunks with respect to the original chunk number
created by the chunk extraction step. The absolute delta between the rows
shows how many chunks are ``lost''\footnote{In the permutation step,
  additional chunks are created.} in each pipeline step.  The tables' last
row ``Type-3-Clone'' reports the total number of simions found after all
steps have been processed. We do not report exact processing times, as the
load of the machines we were using for the analysis varied. Each
individual run for one system took between 1 and 30 hours (depending on
number of chunks).

For the \emph{Info1} data set, we found 105 resp.\ 418 simions. The
higher number for the intent-based strategy is expected, as certain
sub-steps of the implementation can be also seen as individual simions
that are not removed by the subsumption filter if they occur more
often than the surrounding simion. As the data set consists of 109
implementations of the same functionality, we would not expect to find
substantially more simions with any other approach. Hence, the recall
for purely algorithmic code (no I/O, no custom data structures, etc.)
is good.

Overall (not including the \emph{Info1} set), the analysis discovered
153 simions with the method-based chunking strategy and 294 simions
with the intent-based strategy. Compared to the size of the analyzed
systems, this number seems small.  Furthermore, a manual inspection of
a sample of the reported simions revealed that the results still
contain false positives, \ie chunks that a developer would not
consider functionally similar. These were in most cases functions that
produce the same output for a small number of input values. However,
these were mostly ``trivial'' corner cases regarding the input data
(\eg \texttt{null} or the empty String) and the manual inspection of
these functions revealed that they differ significantly in the
behavior for ``interesting'' input data values, thus not qualifying as
simions according to our definition.  Examples for valid simions
included functions that used different data types (such as primitive
types and their corresponding wrapper types) but performed an equal
function. Other valid simions were functions that simply delegated to
another one (in some cases with additional error handling in the
delegating function), thus trivially exhibiting equal functionality.

RQ~2 quantified the numbers of chunks that are discarded due to
different technical reasons, such as missing input. A chunk can be
affected, however, by more than one of these issues. To understand how
many chunks are affected by none of these problems, we have to look at
the entire processing pipeline. On average, about 28\% of chunks for
both strategies survive the execution step and all preceding
steps. This is still a significant part of the systems in which we can
in principle find simions using our approach. Yet, this also means
that more than two thirds of the chunks are lost before the actual
comparison can be performed and is thus not receptive for our dynamic
detection.

Another observation is that while the intent-based chunking strategy
results in the higher number of simions found, the relative number of
simions found in comparison to the number of input chunks is four
times higher for the method-based strategy. One explanation is that
the method boundary chosen by the developer is much more likely to
encapsulate a reusable fragment. Fragments from within a method are
more likely to not be sensible for another developer, thus the
probability of finding a duplicate of it is lower. 
This means that the intent-based strategy is preferred in
terms of results (more simions), but compared to the required
computation time (which scales linear in the number of chunks) the
method-based approach is more effective.

The tables also hint at the amount of cloning. In the method-based case on
average 27\% of the chunks are discarded early on as type-1.5 clones (for
the intent-based strategy this number is lower with about 4\%). From the
simions found, 41\% (or 22\% for intent-based) could be found by a type-3
clone detector. The absolute numbers provide an even clearer picture. There
are 17 times as many chunks removed by the type-1.5 clone filter as there
are simion instances reported (9 times for intent-based), so the cloning
problem seems to be worse than the problem of independent reimplementation
(simion).  Additionally, clones (even of type-3) are much easier to detect
than simions (both in terms of the involved algorithms and the required
processing time) and the detection of clones is also possible for code
where we can not find suitable input or a compilation context.  From a
quality improvement standpoint, this indicates that detecting simions
should only be performed, after all clones (the ``low hanging fruits'')
have been filtered.

\section{Threats to Validity}

One possible threat to the validity of our results are errors in our
implementation of the detection pipeline. To mitigate this, we integrated
excessive logging in our tool and inspected samples of the reported or
excluded chunk at every pipeline step during development. Additionally, we
included the study objects from~\cite{juergens2010code}, for which the
number of simions is known, into our study objects. This ensures that our
approach and implementation are capable of finding at least certain
simions.

To improve the validity of the results we chose study objects we were
familiar with from earlier experiments in a code analysis and quality
context. This helped us to interpret the results compared to an entirely
unknown system. Still, we attempted to select study objects of different
types and sizes, to improve transferability of our results to other Java
systems.

An internal threat to validity is that the different filters used can
distort the results for individual technical challenges. We mitigated
this threat by separate configurations for collecting different
data. The configuration for the results of RQ~2 uses less filters so
we could get the complete results.

\section{Related Work}
Work related to ours can be found in the area of clone detection (for
an overview,
see~\cite{koschke-2007_KoschkeR-clone_survey,roy-2007-clone_survey}),
where syntactically similar code is searched. As the techniques used
there only work on a textual semi-structured representation, however,
they cannot be used to find code fragments that are semantically
similar but not syntactically similar as shown
in~\cite{juergens2010code}. Another related area is equivalence
checking (\eg~\cite{cousineau-1979,raoult-1980,bertran-2005}), which
is a well-studied but undecidable problem. While these papers provide
a theoretical foundation, they do not provide techniques for finding
equivalent code in large real-world systems.  We also use techniques
known from random testing, such as \cite{pacheo-2007-randoop}, but our
setting is different as we require the same input for many test cases
(chunks).

\subsection{Comparison to Jiang\&Su}
To the best of our knowledge, the work of 
Jiang\&Su~\cite{jiang2009automatic} is the only published attempt of automatically
mining semantically equivalent code from a large code base. Thus, we
compare our approach and results in detail to their paper in this
section. The most obvious difference is that our tool chain works on Java
code, while their tool, called \emph{EqMiner}, deals with the C
language. For a more systematic comparison, we structure the comparison
according the different phases of the detection pipeline.

\myparagraph{Chunk Extraction Phase}
This phase is called \emph{code chopper} in EqMiner. Their approach
corresponds to the \emph{sliding window} chunking strategy with a minimal
window size of 10 statements. They report that for long functions the
quadratic number of chunks created this way is too large, which
matches our observation. To mitigate this, EqMiner ``randomly selects up to
100 code fragments from all code fragments in each function''. 
We expect, that this random selection can cause relevant chunks to be missed. Thus, we
employ a strategy based on logical separation found in the syntax, which
also helps to reduce the number of chunks, but hopefully better captures
the programmer's intent compared to random chopping.

\myparagraph{Input Generation Phase}
The type system of C essentially consists of primitive types,
pointers, and record types (structs). Consequently, the input
generator used in EqMiner supports generation of all of these types,
including dynamic allocation of structs to provide a pointer to a
struct, but the generation of arrays is not supported. In Java, there
are also primitive types and arrays, but instead of structured data
(structs) and pointers, Java has classes and object references. While
similar in the intent, this complicates input generation. Furthermore,
the presence of abstract types and interfaces leads to situations
where a suitable concrete implementation can not be easily generated
as input. As long as non-abstract classes are used, we follow the
approach from~\cite{pacheo-2007-randoop} by picking a random
constructor and recursively generating inputs for its parameters.

\myparagraph{Chunk Execution Phase}
This phase in our tool corresponds to the \emph{code transformation} and
\emph{code execution} steps in EqMiner. The obvious difference is that to
make a fragment of Java code compile requires slightly different
surrounding code than with C code. Especially the access to local
attributes and methods within the same class requires additional
considerations. The main difference, however, is how we treat function
calls within the extracted chunk. In EqMiner, Jiang\&Su ``view each
callee as a random value generator and ignore its side-effects besides
assignments through its return values (\ie the random values)''. We found
this approach too limiting for Java code, as we might miss
interdependencies between method calls (for example code might rely on
getting the same value from a getter method that was earlier passed to the
corresponding setter method). Instead we just execute the original method,
which means that the entire context of the chunk must be reconstructed,
including the surrounding class with its attributes and methods.  Actually
executing the methods also requires protection of the execution environment
of unwanted side-effects. For example, feeding code that deletes files with
random data could cause problems during analysis. Using Java's security
manager, however, we could ensure the program to not cause these kinds of
problems.

\myparagraph{Result Comparison Phase}
EqMiner treats  code chunks as equivalent, if they produce the same output
for ten different random inputs. Jiang\&Su also report, that a common
pattern found in the largest clusters of equivalent code is that the
(single) output of the chunk is exactly the input value. So, the chunk is
essentially equivalent to the identity (or a projection of inputs). Our
experiments also showed this pattern and manual inspection of these chunks
revealed that these fragments typically influence the system by other means
(for example by calling functions with side-effects) and would not be considered
equivalent by a developer. We filtered out all chunks following this pattern
prior to comparison. Our tool discards chunks that return the same result
for all random inputs. As the results for RQ~3
indicate, both cases are frequent.

\myparagraph{Study Objects}
Jiang\&Su evaluated their tool on a sorting benchmark and the Linux
kernel. Both systems do not contain code that performs I/O operations
(actually the kernel \emph{offers} system calls for performing I/O) or deal
with UIs. Additionally, a huge part of the kernel deals with process
scheduling, device drivers, or memory management, which are all not known
to require complex string processing. Contrary, as shown by RQ~2, our study
objects spend lots of code on I/O and UI tasks, and string processing is
essential for parts of them. As I/O and UI are tricky for input generation,
and string algorithms are often hard to differentiate with only 10 inputs,
the number of valid chunks and false positives is affected by our choice of
systems (which are typical for Java programs).

\section{Discussion} \label{sec:discussion}

The low number of simions raises the question whether there are no simions
in those systems or rather our detection approach is flawed. Ideally, we
would answer this question by calculating the recall, \ie the fraction of
all known simions we are able to detect. For realistic systems, such as our
study objects, the number of existing simions is not known (and practically
infeasible to determine manually by inspection for even a small part of
them). For artificial benchmarks, such as the Info1 set, we have a good
estimate of the number of simions. Yet, while our recall is good for
this study object, the comparison with the numbers from
Tables~\ref{tab:overal_mb} and~\ref{tab:overal_ib} shows that they are not
representative for realistic systems.

Intuition and experience tells us, that developers tend to reinvent
the wheel and so we would expect many simions.  One explanation for
the low rates could be that we only analyzed simions within a single
system.  Maybe developers know their \emph{own} code base well enough
to reuse code (either by referencing or cloning it) instead of
reimplementing it as a simion. First experiments with detection of
simions between two independent projects, however, did not reveal
substantially higher simion rates. One explanation is given by
Table~\ref{tab:rq2}, which reports high rates of chunks with project
specific data types. As we only find simions with the same I/O
signature, these chunks can not be cross-project simions.

In our opinion another reason for the low detection rates is that the
notion of I/O equivalence is inappropriate. Often, code encountered in
practice might be intuitively counted as a simion, but does not
exhibit identical I/O behavior. Reasons are differences in special
cases, the error handling, or the kind of data types used at the
interfaces. An extreme example would be databases from different
vendors. While they basically provide the same functionality and
interface (SQL), migrating from one database to another is typically
far from trivial as the behavior is \emph{not} the same in all
details. Thus, we believe that there should be a better definition of
a simion than the I/O behavior of code chunks. Finding a suitable
definition and exploiting it for simion detection is one of the main
open questions for future work in this area.

\section{Conclusions and Future Work}

In this paper, we presented an approach for detecting functionally
similar code fragments in Java systems, which was inspired by an
existing approach for C systems. We evaluated the approach for 5 open
source systems and an artificial system with independent
implementations of the same specification. In contrast to existing
work targeting C systems, we experienced low detection results.  In
our opinion, this is mainly due to the limited capability of the
random testing approach. In many cases, input generation either fails
to generate valid input or the generated input is not able to achieve
sufficient code coverage. There is also reason to believe that
similarities are missed due to the chunking, \eg if code fragments
perform a similar computation but use different data structures at
their interfaces. Further research is required to quantify these
issues. To support further investigation by other researchers, we
provide the implementation of our approach for
download\footnote{\url{http://www4.in.tum.de/~ccsm/simions}}.

As future work we plan to manually assess the detected simions in more
detail and check whether they really represent simions according to
our definition. Furthermore, we intend to apply the approach to detect
simions \emph{between} projects. Another interesting idea is to employ
advanced test generation methods, \eg feedback-directed random
testing, or white-box techniques, to achieve higher coverage rates.
We also want to reconsider the definition of simions to better reflect
the intuitive notion of similar functionality. A different definition
might inspire new ways of detection with better results.

\subsection*{Acknowledgments}

We thank the Google Research Awards Program for awarding our
proposal ``Detecting Behavioral Redundancy and Clones in Software
Source Code''.


\bibliographystyle{IEEEtran}
\bibliography{csmr2012_simions}

\copyright 2012 IEEE. Personal use of this material is permitted. Permission from IEEE must be obtained for all other users, including reprinting/republishing this material for advertising or promotional purposes, creating new collective works for resale or redistribution to servers or lists, or reuse of any copyrighted components of this work in other works.

\end{document}